\documentstyle[12pt]{article}

\begin{document}
\baselineskip 22pt

\vspace{1.5cm}
\begin{center}
\begin{Large}
\begin{bf}
Massive Vector Gauge Theory \\ and Comparison with Higgs-Connes-Lott   \\
\vspace{15mm}
\end{bf}
\end{Large}

\vspace{2mm}

{\large Chang-Yeong  Lee${}^{*}$ and Yuval Ne'eman${}^{\sharp}$} \\

\vspace{5mm}

 *: \ {\it Department of Physics, Sejong University \\ Seoul
    143-747, Korea }\\
 $\sharp$: \ {\it Sackler Faculty of Exact Sciences,
     Tel-Aviv University\\  Tel-Aviv 69978, Israel}\\
 
\vspace{15mm}
{\large {\bf Abstract}}\\
\end{center}

\noindent
A massive vector gauge theory constructed from the 
matrix derivative approach of noncommutative geometry is compared 
with the Higgs-Connes-Lott theory. 
In the massive vector gauge theory, 
a new extra shift-like symmetry which is due to the
one form constant matrix derivative allows the theory to have a mass 
term while keeping the gauge symmetry intact.
In the Higgs-Connes-Lott theory, 
the transformation of scalar field makes up 
the deficiency of symmetry
due to the mass term.
Thus, when the scalar field is absent there remains no gauge  
symmetry just like the Proca model.
In the massive vector gauge theory, the shift-like symmetry 
makes up the deficiency of symmetry due to the mass term
even in the absence of the scalar field.
\\

\thispagestyle{empty}
\vfill


\noindent
\hbox to 10cm{\hrulefill}\\
\baselineskip 12pt
\indent
*: leecy@phy.sejong.ac.kr
\\
\indent 
$\sharp$: Also on leave 
 from Center for Particle Physics, 
University of Texas, Austin, Tx 78712, USA \\
\pagebreak

\baselineskip 22pt

\noindent
{\large {\bf I. Introduction}}\\

So far, the Higgs mechanism \cite{higgs} is the only known
 viable mass 
generating mechanism for gauge bosons in four dimensions.
In 1990, Connes and Lott \cite{conlot}
showed that the standard model can be obtained from
the framework of noncommutative geometry. In that framework,
the scalar field is treated as an element of the connection
in the enlarged noncommutative space and thus the Yang-Mills-Higgs
action is naturally obtained from a noncommutative version of
the Yang-Mills action. And the ad hoc insertion of the negative mass
squared term for spontaneous symmetry breaking is replaced by
a generalization of the Dirac operator.
In this noncommutative framework, 
symmetry breaking does not occur in a spontaneous manner, rather
it is built in from the beginning by the discrete part of the
 generalized Dirac operator. 
However, both Higgs mechanism and Connes-Lott model
contain one form gauge field and 
zero form scalar field, and their field contents 
and physical implications are the same.
So, we will call these two Higgs-Connes-Lott theory from now on.

Recently, in Ref. \cite{lee} it was shown that one can construct a
massive gauge theory without introducing a scalar field by generalizing
the matrix derivative approach \cite{sch} of noncommutative geometry.
In this construction, the gauge field has an extra shift-like symmetry
besides the usual gauge symmetry, and that is due to the generalized
matrix derivative consisting of a constant one form even matrix.
In the ordinary gauge theory, if one adds a mass term to the Yang-Mills
action, then gauge symmetry is broken and there remains no gauge symmetry.
This is the case of the massive Proca model.
However, in the massive gauge theory, the deficiency of symmetry
due to the mass term is cured by the extra shift-like symmetry of the
gauge field. And this extra symmetry
 exactly does the role of the scalar field in the
Higgs-Connes-Lott theory, in which the scalar field
 transformation compensates the deficiency
of symmetry due to the mass term
 and thus makes the underlying gauge symmetry maintained.

In this paper, we first briefly review the Higgs-Connes-Lott theory in 
Section II, then in Section III
 we explain the massive vector gauge theory constructed
in the framework of the
 matrix derivative approach of noncommutative geometry.
In Section IV, we compare the symmetry properties of the Higgs-Connes-Lott
theory and
the massive vector gauge theory. 
We conclude in Section V.
\\

\noindent
{\large {\bf II. Higgs-Connes-Lott theory}}\\

In 1995, Connes \cite{conn} has
 modified the Connes-Lott model \cite{conlot}
slightly based on the reality of the spectral triple 
$ \{ {\cal A ~ H ~ D} \}.$
Here, we shall follow this revised version, 
and its technical details will be refered to Ref. \cite{conn}.

A spectral triple is given by an involutive algebra of operators 
${\cal A }$ in a Hilbert space ${\cal H}$ and a selfadjoint operator
${\cal D}$ in ${\cal H}$. It is called even when there exists a
$Z_2$ grading operator $\Gamma$ in ${\cal H}$, 
and otherwise it is called odd.
Here, $\Gamma$ commutes with any element in ${\cal A }$, and anticommutes
with ${\cal D}$.

The algebra ${\cal A }$ plays the role of the algebra of coordinates 
on the space, and it can be noncommutative, and
thus the name noncommutative geometry.
${\cal D}$ is usually called the generalized Dirac operator, since in 
the Riemannian case it just becomes the usual Dirac operator.
 $\Gamma$ corresponds to the chirality operator in a $Z_2$ graded 
Hilbert space, and in some sense
this is related to the orientation of the space that one deals with.

Another important ingredient of the Connes-Lott approach is the
$\pi$ representation of the universal differential
envelop of ${\cal A}$, $\Omega^{*}({\cal A})$, where
$\Omega^{*}({\cal A}) = \oplus \Omega^{k}({\cal A})$ such that
$\Omega^{0}({\cal A})={\cal A}$ and
$\Omega^{k}({\cal A})= \{ a_{0}\delta a_{1} \cdots \delta a_{k}; \;
 a_{0}, a_{1}, \cdots, a_{k} \in {\cal A} \}$, the space of universal
k-forms.
The differential $\delta$ satisfies
$ \delta^2 =0, \; \; \delta(a_{0}\delta a_{1} \cdots \delta a_{k})=
\delta a_{0}\delta a_{1} \cdots \delta a_{k} \in \Omega^{k+1}({\cal A}),$
and the involution * is given by
$(a_{0}\delta a_{1} \cdots \delta a_{k})^* =\delta a_{k}^* \cdots
\delta a_{1}^* a_{0}^* .$
Now, $\pi$ is a map from  $\Omega^{*}({\cal A})$ to ${\cal B (H)}$, the
space of bounded operators on ${\cal H}$, given by
\begin{equation}
 \pi(a_{0}\delta a_{1} \cdots \delta a_{k})=\rho(a_{0})
[{\cal D}, \rho(a_{1})] \cdots [{\cal D}, \rho(a_{k})] ,
\end{equation}
where
$\rho$ is a faithful representation of ${\cal A}$ by bounded
operators on the Hilbert space ${\cal H}$. 
Notice that, in order to respect the nilpotency of $\delta$, ${\cal D}$
should satisfy $ [{\cal D}, [{\cal D}, \ \cdot \ ] ] =0  $ in this 
representation.

A tensor product of two noncommutative spaces with spectral triples, 
 $({\cal A}_{1},{\cal H}_{1},{\cal D}_{1},\Gamma_{1})$ and
$({\cal A}_{2},{\cal H}_{2},{\cal D}_{2},\Gamma_{2})$, is defined as
\begin{eqnarray}
 {\cal A}&=&{\cal A}_{1}\otimes {\cal A}_{2}, \; \;
 \; \; \;   {\cal H}={\cal H}_{1}\otimes {\cal H}_{2}, \nonumber \\
{\cal D}&=&{\cal D}_{1} \otimes 1 + \Gamma_{1}\otimes {\cal D}_{2} ,
     \label{c2} \\
   \Gamma &=&\Gamma_{1}\otimes \Gamma_{2}, \; \; \; \; \;
  \Omega^{*}({\cal A})=\Omega^{*}({\cal A}_{1}) \otimes
    \Omega^{*}({\cal A}_{2}). \nonumber
\end{eqnarray}
Now, we briefly review some simple models in the Connes-Lott approach.
\\

\noindent
{\bf 1. Two-point space} \\

We take ${\cal A}= {\bf C} \oplus {\bf C},$
${\cal H}= {\bf C}^N \oplus {\bf C}^N,$ and
${\cal D}= \left( \begin{array}{cc} 0 & M \\
 \bar{M}  & 0  \end{array} \right),$  $\Gamma=\left( \begin{array}{cc}
 1 & 0 \\ 0  & -1  \end{array} \right)$ where $M$ is an $N\times N$
matrix. Assume $a=(\lambda, \lambda') \in {\cal A}$, then
$\pi(\delta a)= [{\cal D}, \rho(a)]= (\lambda -\lambda')
\left( \begin{array}{cc} 0 & - M \\
 \bar{M}  & 0  \end{array} \right)$.
The connection is given by
${\cal J} = a_{0} \delta a_{1}$ with $ a_{0}=(u,u'), \;
a_{1}=(v,v') \in {\cal A}$, and thus
 $\pi({\cal J})=\pi(a_{0} \delta a_{1})=
\rho(a_{0})[{\cal D}, \rho(a_{1})]=(v-v') \left( \begin{array}{cc}
0 & -u M \\ u' \bar{M}  & 0  \end{array} \right).$ Also,
from
$ {\cal J} = a_{0} \delta a_{1}=(u,u')\cdot (v'-v,v-v')
=(u(v'-v),u'(v-v'))$ and denoting it by ${\cal J}\equiv
(\phi , \bar{\phi} )$, we can write $\pi({\cal J})=\left( \begin{array}{cc}
0 & \phi  M \\ \bar{\phi}  \bar{M}  & 0  \end{array} \right)$.
The $\pi$ representation of the curvature $\theta$ is given by
$\pi(\theta)=\pi(\delta {\cal J} + {\cal J}^2)=(\vert \phi +1 \vert^2
-1) \left( \begin{array}{cc}
M \bar{M}  & 0 \\ 0  & \bar{M} M    \end{array} \right).$
Thus, the Yang-Mills action is given by
$I_{\bigtriangledown}={\rm Tr}_{\omega}((\pi(\theta))^2
{\cal D}_{\bigtriangledown}^{-n})$ where ${\rm Tr}_{\omega}$ is
the Dixmier trace, $n$ is the dimension of the manifold, and
${\cal D}_{\bigtriangledown}$ is the ``covariant derivative" given by
${\cal D}_{\bigtriangledown}={\cal D} + \pi({\cal J})$.
Since $n$ is zero and the Dixmier trace becomes the
usual trace in the present case, the Yang-Mills action is given by
$I_{\bigtriangledown}=2 (\vert \phi +1 \vert^2 -1)^2
{\rm Tr} (M \bar{M})^2 .$ This is just the Higgs potential with minima
at $\phi =0, -2$, and its type indicates
explicitly broken symmetry.
\\

\noindent
{\bf 2. Spinmanifold } \\

Take  ${\cal A}= C^{\infty}(Z) \otimes {\bf C}$ where $Z$ is a 4-dimensional
spinmanifold, and let ${\cal H}= L^2(S)$ where $S$ is the vector bundle of
spinors on $Z$. Then, the Dirac and the chirality operators become the usual
ones, ${\cal D}= \gamma^{\mu}\partial_{\mu}$ and $\Gamma=\gamma_{5}$.
The connection is an ordinary differential 1-form on $Z$, $ \pi({\cal J})=A
=A_{\mu}dx^{\mu}$.
Thus, the curvature is given as the usual one $\pi(\theta)=F=dA +A^2$,
and the Yang-Mills action is
$I_{\bigtriangledown}=\int_{Z} {\rm Tr}(F*F)$.
\\

\noindent
{\bf 3. U(1)xSU(2) model} \\

Take  ${\cal A}= C^{\infty}(Z) \otimes {\cal A}_F$ where
${\cal A}_F = {\bf C} \oplus {\bf H} $ and  
$ {\cal H}= L^2(S)\otimes {\cal H}_F $ where 
$ {\cal H}_F ={\cal E} \oplus \bar{\cal E}$.
Here, we will consider the lepton part only, and ${\cal E}$ is the 
finite dimensional Hilbert space whose basis is labeled by all
leptons and $\bar{\cal E}$ denotes its complex conjugate
Hilbert space. ${\cal D}$ is given by 
\[ {\cal D} =  \gamma^{\mu}\partial_{\mu} \otimes I +
   \gamma_{5} \otimes \left( \begin{array}{cc} Y & 0 \\
 0  & \bar{Y}  \end{array} \right), \]
where $Y=  \left( \begin{array}{cc} 0 & M \\
 \bar{M}  & 0  \end{array} \right)$ and the lepton mass matrix 
$M$ is given by $\left( \begin{array}{cc} 0  \\
 1  \end{array} \right) \otimes {\rm diag}(m_e, m_{\mu}, m_{\tau}).$
The real structure $J$ is given by the charge conjugation times 
$J_F$ where 
\[ J_F(\xi, \bar{\eta}) = (\eta, \bar{\xi}) ~~~~  \forall
 ~  \xi \in {\cal E},
 ~~~ \bar{\eta} \in \bar{\cal E} . \]
The algebra ${\cal A}$ acts on ${\cal E}$ and $ \bar{\cal E}$ 
in the following manner.
Let $a=(\lambda,q) \in {\bf C} \oplus {\bf H}$,
then its action on ${\cal E}$ is given by the following.
Only $q$ acts on lepton doublets, 
$ a \left( \begin{array}{c} \nu_e \\
 e  \end{array} \right)_L  = 
q \left( \begin{array}{c} \nu_e \\
 e \end{array} \right)_L $, and only $\lambda$
acts on lepton singlets, 
 $a e_R = \bar{\lambda} e_R$.
On $ \bar{\cal E}$,
only $\lambda$ acts
\[ \lambda \bar{\xi} = \overline{ (\bar{\lambda} \xi)}, ~~~~~ \forall ~ 
\lambda \in {\bf C}, ~~  \xi \in {\cal E}, ~~ 
\bar{\xi} \in \bar{\cal E}, \]
 and $q$ has no action. 
Note that the action of $J a^* J^{-1}$ on the lepton doublet in ${\cal E}$
is given by multiplication by $\lambda$, and the same action on 
$\bar{\cal E}$ makes $q$ act on the lepton doublet.

If we evaluate the connection and curvature and construct the 
Yang-Mills action following the two-point space case that we did
above, we obtain the
 following action of the
Yang-Mills-Higgs type after some calculation.
\begin{eqnarray}
I_{\bigtriangledown} & = & \int d^4 x {\rm Tr} \{ -\frac{1}{4}
 (F_{\mu \nu}^{B})^2 - \frac{1}{4} (F_{\mu \nu}^{W})^2 
     \label{ymh} \\
 &  &   +
 \sum_{leptons} [ \bar{f}_L \gamma^{\mu}(\partial_{\mu} + B_{\mu}
+ W_{\mu}) f_L   
    +  \bar{f}_R 
 \gamma^{\mu}(\partial_{\mu} + B_{\mu}) f_R ] \nonumber \\
 & &   + \sum_{f f'}
 [M_{f f'} \bar{f}_L \phi f'_R + {\rm C.C.}]  
   + | ( \partial_{\mu} + B_{\mu}
+  W_{\mu}) \phi |^2 - \alpha |\phi|^2 + \beta (|\phi|^2)^2 
 \} \nonumber
\end{eqnarray}
Here, $\alpha$ and $\beta$ are positive real parameters,  
  $B$ and $W$ are U(1) and SU(2) gauge fields, respectively, 
 $\phi$ is a complex scalar doublet, and $f_L$ and $f_R$ are
lepton doublets and singlets, respectively.  
\\

\noindent
{\large {\bf III. Massive vector gauge theory }}\\

In 1990, Ne'eman and Sternberg \cite{ns} first applied the concept of
superconnection \cite{qui}  for the Higgs mechanism.
They considered 0-form scalar field and 1-form gauge field as a multiplet
of a superconnection and wrote down the Yang-Mills-Higgs action 
by inserting the negative mass squared term for the scalar field.

With the concept of matrix derivative, this was done more naturally 
without inserting the negative mass squared term  
in the noncommutative framework \cite{sch,couqet,lhn}.
In this section, 
we first consider how one can generalize the matrix derivative
 from the superconnection viewpoint \cite{lee2}, 
then we construct the massive vector gauge
theory using the generalized matrix derivative \cite{lee}. 
 
Consider a super (or $Z_{2}$-graded) complex vector
space, $V = V^{+} \oplus V^{-}$. 
The algebra of endomorphisms of $V$ is a superalgebra with
the even or odd endomorphisms.
Let ${\cal E}={\cal E}^{+} \oplus {\cal E}^{-}$ be
a super (or $Z_{2}$-graded) vector bundle over a manifold $M$, and
$\Omega(M) =\oplus \Omega^k(M)$ be the algebra of smooth differential
forms with complex coefficients.
Then, the space of ${\cal E}$ valued
differential foms on $M$, $\Omega(M, {\cal E})$,
 has a $Z \times Z_{2}$ grading, and the fibers of ${\cal E}$ are
superspaces. 
Here we are
mainly concerned with its total $Z_{2}$ grading. 

In Ref. \cite{lee2}, we showed that a generalization of the superconnection
concept can yield the matrix derivative of the noncommutative
geometric gauge theory \cite{sch}.
There, the generalized superconnection 
is given by \cite{lee2}
\begin{equation}
 \bigtriangledown =  {\bf d}_{t} + \omega , 
\end{equation}
where
 $ \; {\bf d}_{t}= {\bf d}+{\bf d}_{M}$ is a generalization of the
one form exterior derivative satisfying the derivation property, 
and  $\omega$ is a generalized connection given by
$\omega = \left( \begin{array}{cc} \omega_{0} & L_{01} \\
  L_{10}  & \omega_{1}  \end{array} \right).$ Here,
$\omega_{0}, \ \omega_{1}$ are matrices of odd degree differential forms
and  $ L_{01}, \  L_{10}$ are matrices of even degree differential forms.
The multiplication rule is given by
\begin{equation}
 (u \otimes a)\cdot (v \otimes b) = (-1)^{\vert a \vert  \vert v \vert}
 (uv)\otimes (ab), \; \; \; u,v \in \Omega(M), \; \; a,b \in {\cal A},
\end{equation}
where ${\cal A}$ is the endomorphisms of $V$.
Then the tensor product of $\Omega(M)$ and ${\cal A}$ belongs to 
the endomorphisms of $\Omega(M, {\cal E})$.
In the matrix representation,  ${\bf d} =\left( 
\begin{array}{cc} d & 0 \\ 0  & d  \end{array} \right) $ 
where  $d$ inside
the matrix denotes the usual 1-form exterior derivative 
times a unit matrix, and 
${\bf d}_{M}$ is given below.
Since ${\bf d}_{M}$ should behave as a part of the superconnection operator
\cite{bgv} in a sense, we write it as a (graded) commutator operator
\begin{equation}
{\bf d}_{M}= \left[ \eta, \; \cdot \ \right], \; \; \; \eta \in
   \Omega (M, {\cal E}).
\end{equation}
Then,  ${\bf d}_{M}$ should satisfy  
\begin{eqnarray}
  \; \; {\bf d}_{M}^2 &=& 0, \; \; \; \; \; \; \;
   {\bf d}{\bf d}_{M} + {\bf d}_{M}{\bf d}=0, \label{c11} \\
 {\bf d}_{M}(\alpha \beta)& =& ({\bf d}_{M} \alpha) \beta +
(-1)^{\vert \alpha \vert} \alpha ({\bf d}_{M} \beta), \; \; \; \
 \alpha, \beta \in \Omega (M, {\cal E}). \nonumber
\end{eqnarray}
In Ref. \cite{lee}, two simple solutions satisfying
the  above conditions were given by 
\\
 (1)  $ ~ \;  \eta =   \left( \begin{array}{cc} u & 0 \\
 0  & v  \end{array} \right) $ where $u, \; v$ are odd degree closed forms
with their coefficient matrices satisfying $ u^2 = v^2 \propto 1$ or 
$ u^2=v^2=0,$ \\
 (2) $ \; \eta =   \left( \begin{array}{cc} 0 & m \\
 n  & 0  \end{array} \right)$ where $m, \; n$ are even degree closed forms
with their coefficient matrices satisfying $mn=nm \propto 1$ or 
 $mn=nm=0$. \\
If we take the second solution with 0-form $m, \; n$, then this
choice yields the so-called matrix derivative \cite{sch}
 ${\bf d}_{M}= [\eta , \ \cdot \ ] $
with $ \eta =
 \left( \begin{array}{cc} 0 & \zeta \\ \overline{\zeta}  & 0
\end{array} \right)$ where $  \zeta,  \; \overline{\zeta}$ are
 0-form constant matrices satisfying $  \zeta \overline{\zeta} =
\overline{\zeta} \zeta \propto 1 $.

With the use of the generalized superconnection, the
 curvature is now given by
\begin{equation}
 {\cal F}_{t}=({\bf d}_{t} + \omega )^2 = {\bf d}_{t} \omega 
   + \omega^2 .  
\end{equation}
In this formulation, the Yang-Mills action is given by \cite{lhn} 
\begin{equation}
 I_{YM} =  \int_{M} {\rm Tr} 
  ( {\cal F}_{t}^{\star} \cdot {\cal F}_{t} ) 
\label{c8}
\end{equation}
where $\star$ denotes taking dual for each entries of ${\cal F}_{t}$ 
as well as taking Hermitian conjugate.  
The fermionic action is given by 
\begin{equation}
I_{sp} =\int_{M} \overline{\Psi} \gamma^{\mu}
 ({\bf d}_{t}  + \omega  )_{\mu} \Psi, \; \; \;
\Psi \in V \otimes S
\label{c9}
\end{equation}
where $S$ is a spinor bundle.

Now, we consider the first solution for ${\bf d}_{M}$ given 
above with
$ \eta =  \left( \begin{array}{cc} \sigma  & 0 \\
 0  & \sigma'  \end{array} \right) $
where $\sigma, ~ \sigma'$ are constant 1-form matrices whose 
squares are either proportional to a unit matrix or zero.
For the generalized connection $\omega$, we set
\begin{equation}
 \omega =  \left( \begin{array}{cc} A & 0 \\
 0  & A'  \end{array} \right)
\end{equation}
where $A, \ A' $ consist of one forms only.

For a definite understanding, we consider the case where $A, \ A'$ are
SU(2) valued 1-form fields, and $\sigma$ and $\sigma'$ are
proportional to a SU(2) Pauli matrix, say $ \tau_3$:
\begin{eqnarray}
 & & A = \frac{i}{2} A^a_{\mu} \tau_a dx^{\mu} \equiv A_{\mu} dx^{\mu}, 
~~~~ A'= \frac{i}{2} {A'}^a_{\mu} \tau_a dx^{\mu} \equiv {A'}_{\mu} dx^{\mu},
  \nonumber   \\ 
 & & 
\sigma = \sigma' =  \frac{i}{2} m  \tau_3 n_{\mu} dx^{\mu}
 \equiv \sigma_{\mu} dx^{\mu}. \label{nc9}
\end{eqnarray}
Here $\tau$'s are Pauli matrices, $n_{\mu}$ is a constant
four vector, and
$m$ is a constant parameter.
Throughout the paper, we use the metric $g_{\mu \nu}=(-1,+1,+1,+1)$ and 
$\epsilon_{0123} = +1$, and the wedge product between forms is understood.

The curvature  
\begin{eqnarray}
 {\cal F}_{t} & = & {\bf d}_{t} \omega + \omega^2  \\
      & = & {\bf d}\omega + [ \eta , \omega ]_{\pm} +  \omega^2
 \nonumber
\end{eqnarray}
is now given by
\begin{eqnarray*}
 {\cal F}_{t} & = & \left(
\begin{array}{cc} d & 0 \\ 0  & d  \end{array} \right)
\left( \begin{array}{cc} A & 0 \\
 0  & A'  \end{array} \right) + \left[ 
 \left( \begin{array}{cc} \sigma  & 0 \\
 0  & \sigma'  \end{array} \right) , 
 \left( \begin{array}{cc} A & 0 \\
 0  & A'  \end{array} \right) \right]_{+} + 
 \left( \begin{array}{cc} A & 0 \\
 0  & A'  \end{array} \right) \cdot \left( \begin{array}{cc} A & 0 \\
 0  & A'  \end{array} \right).
\end{eqnarray*}
The first and third terms are the usual ones and the second term is a new
piece due to the matrix derivative which we calculate below.
Since all the odd parts are vanishing, the upper and lower diagonal 
parts do not mix. Hence, we mostly consider the upper part in our 
calculation. 
\\
And, 
\begin{eqnarray} 
 \sigma  A + A \sigma 
 &  = &  \frac{1}{2} \left\{ [\sigma_{\mu}, A_{\nu}] 
   - [ \sigma_{\nu}, A_{\mu} ] \right\} dx^{\mu}dx^{\nu}  \nonumber \\
 & = & - \frac{1}{4} m \ n_{ [ \mu} \left( \begin{array}{cc} 0 & A_1 -i A_2 \\
 -A_1 - i A_2  & 0  \end{array} \right)_{\nu ] }  dx^{\mu}dx^{\nu} 
 \nonumber \\
 & \equiv & \frac{1}{2} {\cal A}_{\mu \nu} dx^{\mu}dx^{\nu}. 
\end{eqnarray}
Thus the curvature is given by
\begin{equation}
{\cal F}_{t} = \left( \begin{array}{cc} F_t & 0 \\ 0  & F'_t
  \end{array} \right)
\end{equation}
with
\begin{equation}
 F_t =  \frac{1}{2} (F_{\mu \nu} + {\cal A}_{\mu \nu}) dx^{\mu}dx^{\nu},
\end{equation}  
and $ F'_t$ is the same as $F_t$ except that $A$ is replaced by $A'$,
and $ F_{\mu \nu}$ is the usual one,
\begin{equation}
F_{\mu \nu} =  \partial_{ [ \mu } A_{\nu ] } 
           + [  A_{\mu}  ,  A_{\nu} ]. 
\end{equation}
  Following the same calculational step as in Ref. \cite{lhn}, we 
obtain the Yang-Mills type action of massive gauge fields  
 from Eq. (\ref{c8}), 
\begin{eqnarray}
 I_{YM}& =& 
   \int_{M}{\rm Tr} ( {\cal F}_{t}^{\star} \cdot {\cal F}_{t} ) 
  \label{c16} \\
        & = &   \frac{1}{2} \int_{M} d^4 x \ {\rm Tr} \left[ \left( 
   F_{\mu \nu} + {\cal A}_{\mu \nu} \right) 
   \left( F^{\mu \nu} + {\cal A}^{\mu \nu} \right) +
      \left( {\rm terms ~ with} ~~ A \rightarrow A' \right) \right]
      \nonumber \\
  & = &   \frac{1}{2} \int_{M} d^4 x \ 
        {\rm Tr} \left[  F_{\mu \nu}F^{\mu \nu} +
   F_{\mu \nu}{\cal A}^{\mu \nu} + {\cal A}_{\mu \nu}F^{\mu \nu}
     + {\cal A}_{\mu \nu}{\cal A}^{\mu \nu}
    + \left( {\rm terms ~ with} ~~   A \rightarrow A' \right) \right].
 \nonumber
\end{eqnarray}
The fourth term provides quadratic terms homogeneous in $A_1$
 and $A_2$:
\begin{equation}
   \frac{1}{2} {\rm Tr} {\cal A}_{\mu \nu}{\cal A}^{\mu \nu}  = 
   \frac{1}{2} m^2 \left[ - n_{\mu}n^{\mu} \left( A_{1 \nu} A_{1}^{\nu} 
    + A_{2 \nu} A_{2}^{\nu} \right) +  n_{\mu}n_{\nu} 
           \left( A_{1}^{\mu} A_{1}^{\nu}
    + A_{2}^{\mu} A_{2}^{\nu} \right) \right] . 
\label{c17}
\end{equation}
The second and third terms also give terms quadratic in $A$ but mixed
in  $A_1$ and $A_2$:
\begin{equation}
 \frac{1}{2}{\rm Tr} \left(
 F_{\mu \nu}{\cal A}^{\mu \nu} + {\cal A}_{\mu \nu}F^{\mu \nu}
 \right) =  m \epsilon^{a b} \left( n_{\mu}A_{a \nu} 
               \partial^{\mu} A_b^{\nu} -
            n_{\mu}A_{a \nu}  \partial^{\nu} A_{b}^{\mu} 
         \right) + ~ O(A^3) 
\label{c18} 
\end{equation}
where $a,b=1,2$ and $\epsilon^{1 2} = - \epsilon^{2 1} = 1, 
~ \epsilon^{1 1}= \epsilon^{2 2} =0$.

Before we perform diagonalization and obtain the propagators for these fields,
we first identify the symmetry of the action.
In Ref. \cite{lhn}, the so-called horizontality condition was used
to analyze the BRST symmetry of the noncommutative geometric gauge theory.
Since we use the same superconnection framework, 
the BRST analysis will be more convenient 
for finding the symmetry of
the theory.
In the Yang-Mills theory, the horizontality condtion is given by
\cite{tm,bt,nt,ln}
\begin{equation}
\widetilde{F} \equiv \widetilde{d}\ \widetilde{A}\ + \widetilde{A}\
\widetilde{A}\ =F ,
\label{2c4}
\end{equation}
where
\begin{eqnarray*}
\widetilde{A}\ & =& A_{\mu} dx^{\mu} + A_{N}dy^{N}
+A_{\bar{N}}d {\bar{y}}^{\bar{N}}
 \equiv A + c  + \bar{c}  , \\
\widetilde{d} & =& d + s + \bar{s} , \; d = dx^{\mu}\partial_{\mu}, \;
s = dy^{N}
\partial_{N}, \; \bar{s} =d{\bar{y}}^{\bar{N}}\partial_{\bar{N}}, \\
F & = & dA+AA={1\over 2}F_{\mu\nu}dx^{\mu}dx^{\nu} .
\end{eqnarray*}
Here $ y$ and $\bar{y}\ $ denote the coordinates in the direction of
gauge orbit of the principal fiber whose structure-group is
${\cal G} \otimes {\cal G} $, and $c, \ \bar{c}$ are ghost and antighost
fields.
The above horizontality condition now yields the BRST and anti-BRST
transformation rules for the Yang-Mills theory.
\begin{eqnarray}
 (dx)^1(dy)^{1} & : &
sA_{\mu}=D_{\mu}c
\nonumber , \\
(dx)^1(d{\bar{y}})^{1} & : &
{\bar{s}}A_{\mu}=D_{\mu}{\bar{c}}
\nonumber , \\
(dy)^{2} & : &
s c =-c c
\label{2c5} , \\
(d{\bar{y}})^{2} & : &
{\bar{s}}{\bar{c}}=-{\bar{c}}{\bar{c}}
\nonumber , \\
(dy)^1(d{\bar{y}})^{1} & : &
s{\bar{c}}+{\bar{s}}c =-[ c ,{\bar{c}} ] .
\nonumber
\end{eqnarray}
In the superconnection framework, the horizontality condition is given
as follows \cite{lhn}.
\begin{equation}
  \widetilde{{\cal F}_t} \equiv \widetilde{{\bf d}_t}
\widetilde{\omega} +\widetilde{\omega}  \cdot
\widetilde{\omega} = {\cal F}_t  
\label{c21}
\end{equation}
where
\begin{eqnarray}
 \widetilde{{\bf d}_t} & = &  {\bf d}_{t} + {\bf s} + \bar{\bf s} ,
 \label{c22} \\
\widetilde{\omega} & = & {\omega} + {\cal C} + \bar{\cal C},
\label{c23} 
\end{eqnarray}
and 
\begin{equation}
{\bf s}  =   \left(\matrix{s&0\cr 0&s\cr}\right), ~~
\bar{\bf s}  =  \left( \begin{array}{cc} \bar{s}  & 0 \\
    0 & \bar{s} \end{array} \right), ~~
{\cal C} = \left(
         \begin{array}{cc} c & 0 \\ 0 & c' \end{array} \right),
~~ \bar{\cal C} = \left(
        \begin{array}{cc} {\bar{c}}  & 0 \\ 0 & {\bar{c'}} \end{array}
        \right).
\label{c24}
\end{equation}
The above horizontality condition yields the following BRST and anti-BRST
transformation rules:  
\begin{eqnarray}
(dy)^{1} & : &
{\bf s} {\omega}  =  -{\bf d}_{t} {\cal C}
- {\omega}  \cdot {\cal C} - {\cal C} \cdot  {\omega} , \label{c25}  \\
(d{\bar{y}})^{1} & : &
\bar{\bf s}  {\omega}  =  -{\bf d}_{t}  \bar{{\cal C}}
-  {\omega} \cdot \bar{{\cal C}} -\bar{{\cal C}} \cdot  {\omega},  \\
(dy)^{2} & : &
{\bf s}  {\cal C}  =  -{\cal C} \cdot {\cal C} ,  \label{c26} \\
(d{\bar{y}})^{2} & : & \bar{\bf s} \bar{{\cal C}}   =
-\bar{{\cal C}} \cdot \bar{{\cal C}} ,  \\
(dy)^1(d{\bar{y}})^{1} & : &
{\bf s} \bar{{\cal C}}  + \bar{\bf s} {\cal C}  +{\cal C}
\cdot \bar{{\cal C}}
+ \bar{{\cal C}} \cdot {\cal C}   =  0 .
\end{eqnarray}
Since all the odd parts vanish as before, we again consider only the upper
diagonal (even) parts in our calculation. Then, the BRST
and anti-BRST transformation rules for the fields appearing in the
upper parts can be written as    
\begin{eqnarray}
 & &   s A  =  -d c - [ \sigma , c ]_{+} - [A, c]_{+} , \label{c30} \\ 
 &  & \bar{s} A  =  - d  \bar{c}  - [ \sigma ,\bar{c} ]_{+} 
      -[A, \bar{c} ]_{+} ,  \\
 & &  s  c  =   -c c ,  \label{c32} \\
 & &  \bar{s} \bar{c}   =  -\bar{c} \bar{c} ,  \\
 &  &  s \bar{c}  + \bar{s} c  + c \bar{c} + \bar{c} c   =  0 , 
\end{eqnarray}
where
\[ c=  \frac{i}{2} c_a \tau^{a}, ~~ 
   \bar{c} = \frac{i}{2} \bar{c}_a \tau^{a}, ~~ a=1,2,3 . \] 
Now, one can check that the above BRST and anti-BRST transformations
are nilpotent, $ s^2 = \bar{s}^2 = 0$, and the total curvature 
$F_t=d A + A A + \sigma A + A \sigma $ transforms as the usual
curvature $F=dA + AA$, 
\begin{equation}
s F_t = -[c, F_t].
\label{c35}
\end{equation}
Therefore, our Yang-Mills action, 
\begin{equation}
 I_{YM}^0  =  \int_{M} {\rm Tr} F_t^* F_t  
          =  \frac{1}{2} \int_{M} d^4 x \ {\rm Tr} \left[ \left(
   F_{\mu \nu} + {\cal A}_{\mu \nu} \right)
   \left( F^{\mu \nu} + {\cal A}^{\mu \nu} \right) \right] 
\label{act}
\end{equation}
where $*$ denotes the Hodge dual,
is invariant under the above given BRST(anti-BRST) transformation.
Since the BRST and gauge transformations
for classical fields are the same except for a switch between the
 classical gauge parameter
and the ghost field, one can check that the action (\ref{act})
 is invariant under the following
gauge transformation
\begin{equation}
\delta A_{\mu} = \partial_{\mu} \varepsilon + [ A_{\mu}, \varepsilon] 
    + [\sigma_{\mu}, \varepsilon]
\label{c36}
\end{equation}
where
$\varepsilon = \frac{i}{2} \varepsilon_a \tau^a ~~ (a=1,2,3)$ is a zero 
form gauge parameter.  
In order to obtain the propagators we use the following gauge fixing 
term for the action (\ref{c16})
\begin{equation}
{\cal L}_{g.f.} = \frac{1}{\xi} {\rm Tr} ( {\bf d}_{t} \omega^{\star})^2,
\end{equation}
which is translated into the following condition for the action (\ref{act}), 
\begin{equation}
{\cal L}_{g.f.}^{0} = \frac{1}{\xi} {\rm Tr} (\partial_{\mu} A^{\mu} +
    \sigma_{\mu} A^{\mu} - A_{\mu} \sigma^{\mu} )^2
\label{gfe}
\end{equation}
where $\sigma_{\mu}, ~A_{\mu}$ are given by Eq. (\ref{nc9}).
In terms of 
\begin{equation}
W_{\pm}^{\mu}= \frac{1}{\sqrt{2}}(A_1^{\mu} \mp i A_2^{\mu}),
\label{wdf}
\end{equation}
we obtain the propagators for $W_{\pm}, ~ A_3$, after some calculation:
\begin{eqnarray}
W_{\pm} & : & \bigtriangleup^{\pm}_{\mu, \nu} =
  \frac{1}{(P^{\pm})^2}\left( g_{\mu \nu} + (\xi -1) \frac{P^{\pm}_{\mu}
           P^{\pm}_{\nu}}{(P^{\pm})^2} \right),
\nonumber \\
A_3 & : & \bigtriangleup^{3}_{\mu, \nu} =
   \frac{1}{p^2} \left( g_{\mu \nu} + (\xi -1) \frac{p_{\mu} p_{\nu}}
      {p^2}  
 \right),  \label{prop}
\end{eqnarray}
where $P^{\pm}_{\mu} = p_{\mu} \pm m n_{\mu}$.
If we set $n \cdot p = n_{\mu} p^{\mu} = 0$, then the denominator of the
$W$-propagator becomes $(P^{\pm})^2 = p^2 + m^2 n^2$. 
Thus, choosing $n_{\mu}$ satisfying the above condition exhibits
the $W$ field having a mass; $({\rm mass})^2=m^2 n^2$.

In Eq. (\ref{prop}), the propagators for $W_{\pm}$ look
apparently different by a term involving the gauge fixing parameter $\xi$.
This difference can be removed for the choice of $\xi=1$, and for other
$\xi$ values we expect that
the contribution from this $\xi$ related term be cancelled by
that of ghosts since the gauge fixing should not affect the
physics.

It is also possible to provide mass terms for all the
gauge fields
 by properly choosing a constant one form even matrix for 
the matrix derivative.
For instance, if we replace $\tau_3$ appearing
in $\sigma_{\mu}= \frac{i}{2} m n_{\mu} \tau_3$ with $\tau_1 +
\tau_2 + \tau_3$, then all $A_1, A_2, A_3$ fields become massive.
However, if we use an identity matrix for  $\sigma_{\mu}$, then
there will be no massive vector gauge field.
\\

\noindent
{\large \bf IV. Comparison of symmetries }\\

In the Higgs-Connes-Lott theory, the gauge field transforms in
the usual way
\begin{equation}
 \delta A_{\mu} =  D_{\mu} \epsilon , 
\end{equation}
while the scalar field transforms with an additional term \cite{lhn}
\begin{equation}
 \delta \phi =  \phi \epsilon + \epsilon \zeta 
\end{equation}
where $\epsilon$ is a 0-form gauge parameter and $\zeta$ is an
odd part of a constant 0-form odd matrix which is
 the discrete part of the generalized Dirac operator.
Because of this extra piece in the transformation,
the odd curvature component which provides the mass term for the gauge
boson transforms covariantly,
\begin{equation}
 \delta ( \partial_{\mu} \phi + A_{\mu} \phi + A_{\mu} \zeta ) =
  \epsilon (\partial_{\mu} \phi + A_{\mu} \phi + A_{\mu} \zeta ) . 
\end{equation}
Here, the shift-like part of $\phi$-transformation related to the
discrete part of the Dirac operator cancels the 
non-covariant transformation part of the $\zeta$ term, which
provides the mass term. This makes the 
theory invariant under the gauge transformation even with the mass
term.
However, if there is no scalar field,
 $\phi=0,$ then the action is not invariant under 
$\delta A_{\mu} =  D_{\mu} \epsilon,$ and the action 
exactly resembles the Proca's.

In the massive gauge theory, the gauge field itself has a
shift-like part in its transformation
due to the action of the matrix derivative
\begin{equation}
 \delta A_{\mu} =  D_{\mu} \epsilon + [\sigma_{\mu}, \epsilon] . 
\end{equation}
The shift-like extra piece in this case also does the role of the 
shift-like transformation of the scalar 
field in the Higgs-Connes-Lott case.
And as we have seen in the previous section, the action constructed
with the matrix derivative of even matrix is gauge invariant under
this transformation although it includes mass terms.
This is the way how the gauge theory
 can have a mass term for the gauge field
 while keeping the gauge symmetry intact
even without a scalar field. 

Here, we would like to note a characteristic feature 
of the noncommutative construction.
Because of the shift-like transformation of the scalar field, one 
might wonder whether a simple replacement of $\phi$ with 
$\phi + \zeta$ yields the result of the Higgs-Connes-Lott.
The answer is no. It simply shifts the vacuum by
$- \zeta$, and does not change the potential shape into an inverted
Mexican hat which is 
essential for symmetry breaking. The Connes-Lott model exactly does that
job. The discrete part of the generalized Dirac operator provides
the needed potential shape:
\[ V(\phi) \sim [ (\phi + \zeta )^2 -\zeta^2]^2 . \] 
Also, one can not have the shift-like extra transformation piece for the
scalar field unless the generalized Dirac operator 
has a discrete piece.

In the massive vector gauge theory case, the simple replacement of
$ A_{\mu}$ with $ A_{\mu} + \sigma_{\mu} $ does not do the job either.
As in the Higgs-Connes-Lott case, the extra shift-like piece in the
gauge field transformation, which is due to the matrix derivative
of even matrix, makes this theory have the gauge symmetry
even with the mass term. 
This mechanism exactly parallels to that of the Higgs-Connes-Lott theory,
as the shift-like piece compensates the symmetry breaking piece
of the transformation from the mass term.
\\

\noindent
{\large \bf V. Conclusion }\\

In this paper, we construct a massive vector gauge theory 
which possesses both the usual gauge symmetry and a shift-like
symmetry, then compare it with the Higgs-Connes-Lott theory.

In the usual construction of noncommutative geometric gauge theory,
only a zero form constant odd matrix has been used for the 
matrix derivative or for the discrete part of the 
generalized Dirac operator. 
In the matrix derivative approach,
 this constant zero form odd matrix 
together with scalar fields appearing in the odd
part of the gauge multiplet (or superconnection) give rise to 
the Higgs mechanism. In the Connes-Lott formalism, this constant
odd matrix does the role of the generalized Dirac operator acting
on a discrete space.
However, by constructing the noncommutative geometric gauge theory 
 from the superconnection viewpoint, it is also possible to use 
a constant one form even matrix for the matrix derivative.

The matrix derivative of constant one form even matrix 
provides a shift-like symmetry to the gauge field, and
this shift-like symmetry transformation of gauge field 
does the role of the scalar field transformation in the
Higgs-Connes-Lott theory, where the scalar field
 transformation compensates the deficiency
of symmetry due to the mass term
and makes the underlying gauge symmetry maintained.  
This way it becomes possible to 
construct a massive vector gauge model similar to the Proca's while
keeping its gauge symmetry intact. 
\\

\noindent
{\large \bf Acknowledgments}\\
C. Y. Lee was supported
in part by Korea Ministry of
Education, BSRI-97-2442, and KOSEF grant 971-0201-007-2.
\\

\end{document}